\documentstyle[times,pramana,epsf,floats]{ias}
\input{psfig.sty}
\begin{document}
\mark{{Atmospheric Neutrino Anomaly...}{John M. LoSecco}}
\title{Inconsistencies in Interpreting the Atmospheric Neutrino Anomaly}

\author{John M. LoSecco}
\address{Physics Department, University of Notre Dame, Notre Dame, Indiana
46556, USA}
\abstract{We note a discrepancy between the value of R expected on the basis
of the muon neutrino angular distribution and the value actually observed.
The energy independence of $R$ leads to a fine tuning problem.  This may be
indicative of some unaccounted for new physics}

\maketitle
\section{Introduction}
Evidence for an atmospheric neutrino anomaly is of two types.  The angular
distribution of the SuperKamiokande data shows an upward/downward deficiency.
\begin{table}[h]
\caption{\label{tab:R}Measurements of $R$ for various experiments}
\hskip4pc\vbox{\columnwidth=26pc
\begin{tabular}{|l|c|} \hline
Experiment & Measured R value\\
\hline
IMB-1 & 0.68$\pm$0.11 \\
Kamiokande  Sub-GeV & 0.60$^{+0.06}_{-0.05}$$\pm$0.05\\
Kamiokande  Multi-GeV& 0.57$^{+0.08}_{-0.07}$$\pm$0.07\\
IMB-3 & 0.54$\pm$0.05$\pm$0.12\\
Frejus & 1.00$\pm$0.15$\pm$0.08\\
Nusex & 0.99$\pm$0.29\\
Soudan(99) & 0.64$\pm$0.11$\pm$0.06\\
Super Kamiokande  Sub-GeV & 0.638$\pm$0.016$\pm$0.050\\
Super Kamiokande  Multi-GeV& 0.658$^{+0.030}_{-0.028}$$\pm$0.078\\
\hline
\end{tabular}
}
\end{table}
In addition numbers of prior experiments (table \ref{tab:R}) noted an
apparent deficiency of muon neutrinos.  This deficiency seems to be too large
to be explained (solely) by the angular anomaly.
\begin{figure}[htbp]
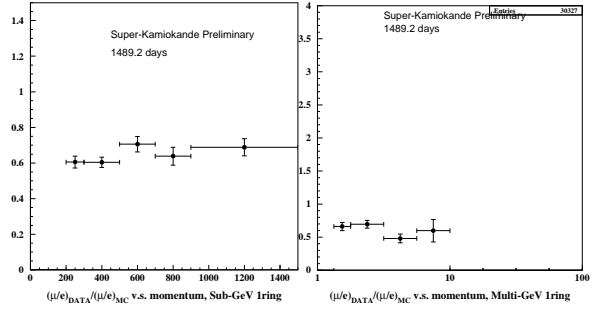

\epsfysize=4cm
\centerline{\epsfbox{sub-gev-r.epsi}
\epsfysize=4cm\epsfbox{multi-gev-r.epsi}}
\caption{R as a function of energy Sub-GeV (left) and Multi-GeV (right)}
\label{fig:sgr}
\end{figure}
\section{$(\mu/e)_{observed}/(\mu/e)_{expected}$}
The quantity $R=(\mu/e)_{observed}/(\mu/e)_{expected}$ was introduced to
minimize systematic errors associated with neutrino flux estimates.
It is insensitive to the flux normalization, insensitive to detection
efficiencies, insensitive to scattering angle errors between the $\nu$ and
the charged lepton (unless binned by direction), insensitive to muon-electron
differences that are modeled in the Monte Carlo.  Table \ref{tab:R} lists
and compares several measurements of $R$.
\begin{figure}[htbp]
\epsfysize=6.5cm
\centerline{\epsfbox{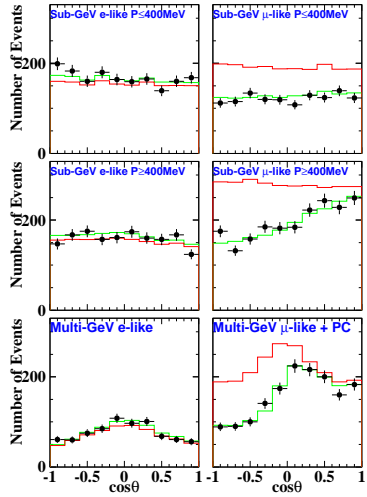}}
\caption{The measured zenith angle distributions}
\label{fig:zendat}
\end{figure}
\vspace{-1cm}
\section{The Angular Distribution}
The flight length of the $\nu$ depends on its direction.\\
$
L(\cos(\theta_{z})) = \sqrt{R_{1}^{2} (\cos^{2}(\theta_{z})-1) + R_{2}^{2}}
-R_{1} \cos(\theta_{z})
$\\
with $R_{2}$ representing the distance from the
center of the earth to the upper atmosphere where the neutrinos are born and
$R_{1}$ representing the distance from the center of the earth to the detector
($R_{2}>R_{1}$).
\begin{figure}[htbp]
\centerline{\psfig{file=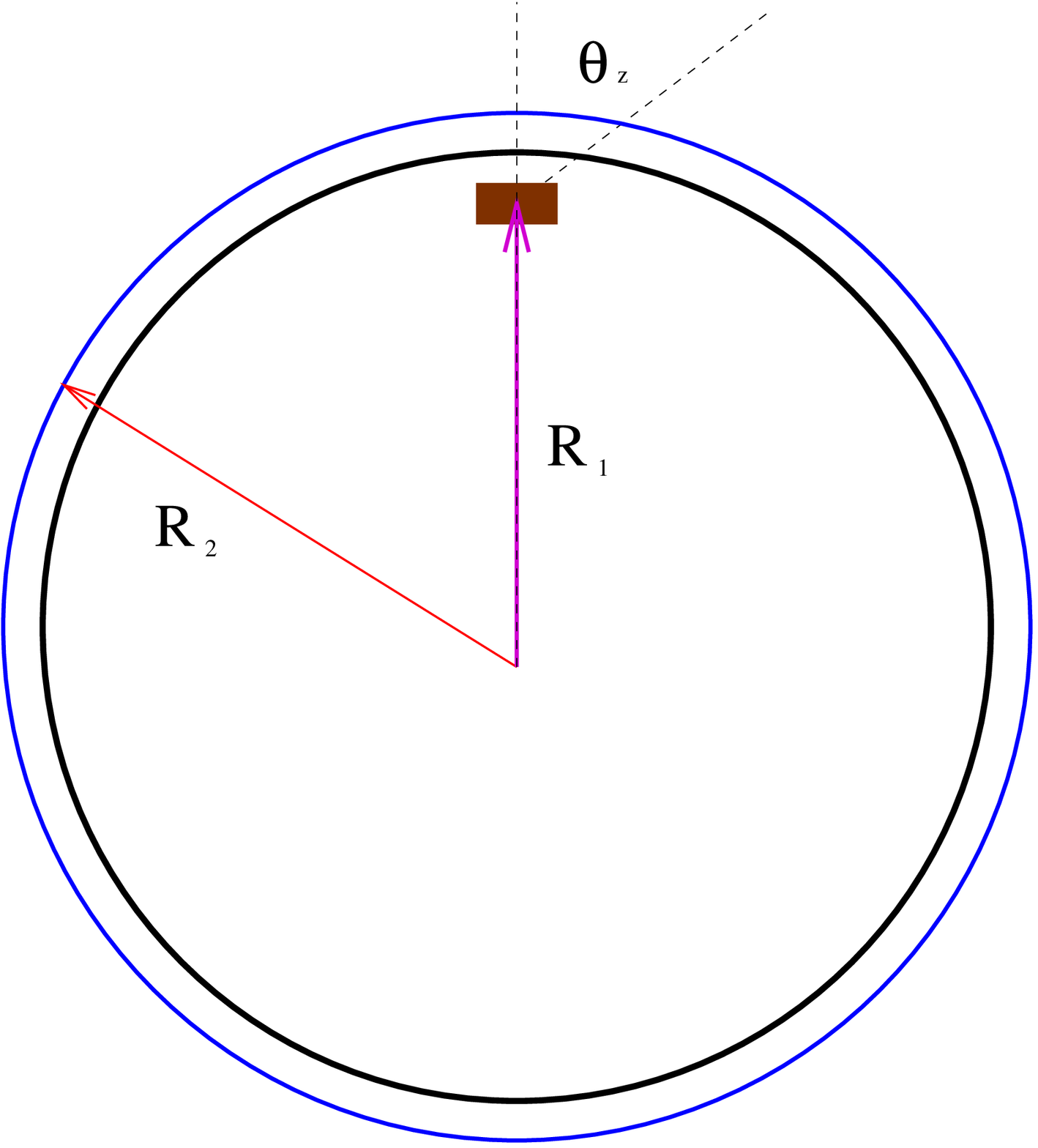,width=5cm,angle=270}
\epsfysize=5cm\epsfbox{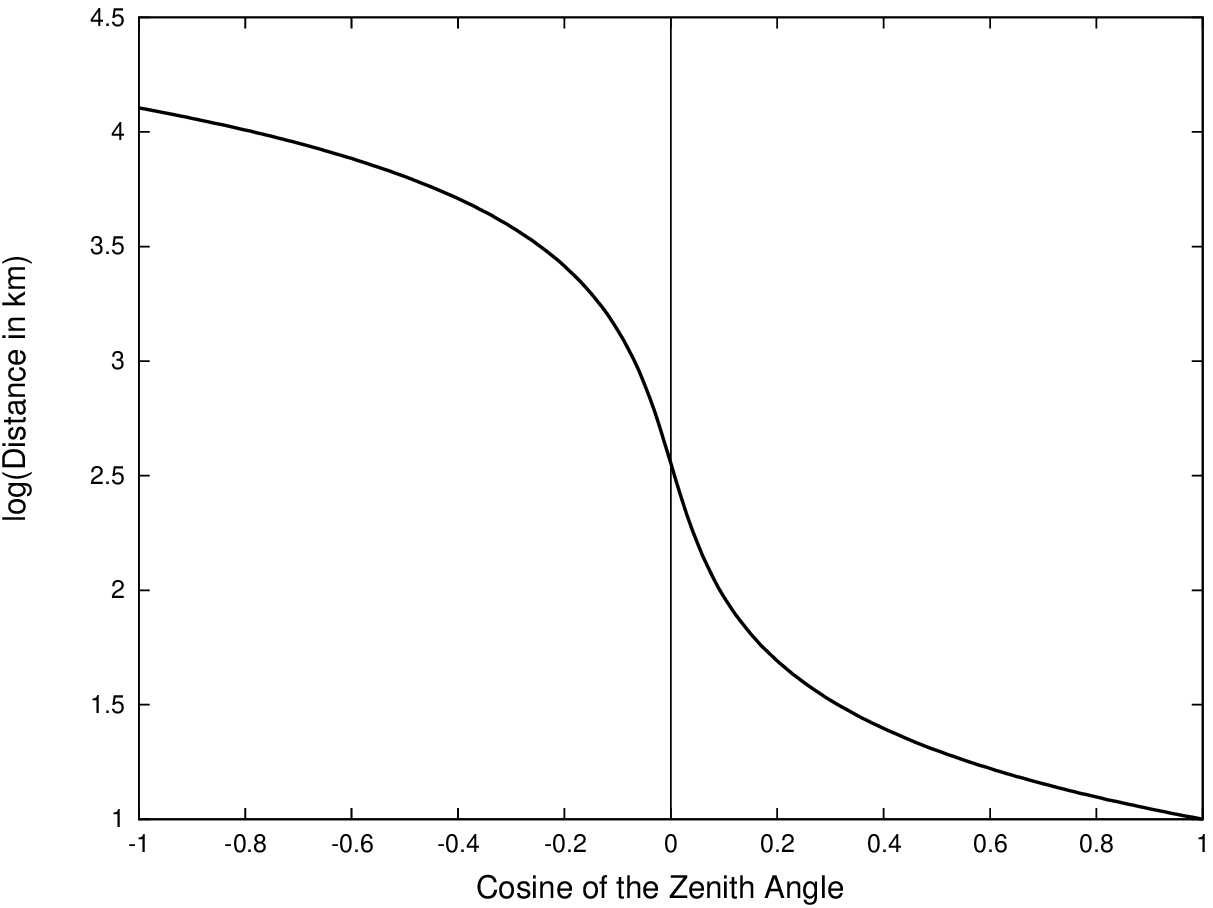}}
\caption{Flight distance as a function of zenith angle}
\label{fig:fdza}
\end{figure}
Those from above travel about 15 km. Those from below travel about 10,000 km.
The propagation distance is a continuous distribution, but it is
predominantly {\em bimodal}.  Such a bimodal distribution facilitates a
comparison of $\nu$ flux traveling over 2 distance scales.  Such a comparison
is insensitive to the flux normalization.
Near the zenith, it is insensitive to first order in the scattering angle
between the reconstructed muon and the neutrino.
\section{The Problem}
The observed muon deficiency is too large to be explained by just
muon neutrino oscillations, even at maximal mixing.  Even if all events in the
upward hemisphere were fully mixed the smallest $R$ could be is about 0.75.
This is a fine tuning problem.
While it is possible that at maximum mixing, for some value of the energy
the integral effect of neutrino oscillations over the set of distances
illustrated in figure \ref{fig:fdza} could produce an $R$ of about 0.63.
A small change in the oscillation length at energies above or below
this tunned value would cause $R$ to jump from 0.75 when only the lower
hemisphere oscillates, at higher energies and to 0.5 when oscillations occur
over all solid angle at lower energies.
But as seen in figure \ref{fig:sgr} $R$ is essentially energy
independent\cite{feb98} with
no clear break over a factor of about 50 in energy.
A more {\em natural} explanation\cite{rofe} would be a large enough
$\Delta m^{2}$ such
that oscillations occur over all solid angle for all energies and
$\sin^{2}(2 \theta) \approx 0.8$.  But this would not lead to any directional
modulation.
The atmospheric neutrino anomaly may be due to new physics perhaps including
neutrino oscillations.
Numerous (2+14) {\em ad hoc} parameters, in addition to neutrino oscillations,
have been used to fit the data in figure \ref{fig:zendat}.
If one attempts to understand the data utilizing
only neutrino oscillations and no adjustments the difference between
the expected and observed values of $R$
With the measured up/down rate the difference is about 4-5 sigma.
Assuming maximal mixing, the difference is about 3 sigma (R=0.75 expected
0.61$\pm$0.05 observed)

\end{document}